\documentclass[prl,twocolumn,showpacs,amsmath,amssymb,superscriptaddress]{revtex4-1}
\pdfoutput=1

\usepackage{graphicx}
\usepackage{epsfig}
\usepackage{dcolumn}
\usepackage{bm}
\begin{document}

\title{Zero field spin polarization in a 2D paramagnetic resonant tunneling diode}

\author{M. R\"{u}th}
\author{C. Gould}
\author{L.W. Molenkamp}

\affiliation{
Physikalisches Institut (EP3) and R\"{o}ntgen Center for Complex Material Systems, Am
Hubland, Universit\"{a}t W\"{u}rzburg, D-97074 W\"{u}rzburg, Germany
}

\date{\today}

\begin{abstract}

We study I-V characteristics of an all-II-VI semiconductor resonant tunneling diode with dilute magnetic impurities in the quantum well layer. Bound magnetic polaron states form in the vicinity of potential fluctuations at the well interface while tunneling electrons traverse these interface quantum dots. The resulting microscopic magnetic order lifts the degeneracy of the resonant tunneling states. Although there is no macroscopic magnetization, the resulting resonant tunneling current is highly spin polarized at zero magnetic field due to the zero field splitting. Detailed modeling demonstrates that the local spin polarization efficiency exceeds 90\% without an external magnetic field.

\end{abstract}

\maketitle

The implementation of device components based on resonant tunneling diodes (RTDs)
is one route towards the elaboration of a full semiconductor spintronics based technology scheme. While a ferromagnet/tunnel barrier spin injector \cite{Crooker2005,Jansen2007,Patel2009} produces a fixed spin polarization for each given magnetization state, dilute magnetic semiconductors (DMS) can be used in II-VI semiconductor RTDs to implement spin selective tunneling at different bias voltages \cite{Slobodskyy2003}. A caveat to this approach has been the paramagnetic nature of bulk (Zn,Mn)Se, which makes the application of an external magnetic field necessary for spin filter operation. This can be overcome by using the 0D states of self assembled quantum dots embedded in a DMS host material, since the microscopic magnetic environment of a dot allows for the formation of bound magnetic polaron (BMP) like states which lift the spin degeneracy for the tunneling electrons \cite{Dietl1982,Gould2006}. Such self assembled quantum dot structures have a rich resonance spectrum which typically occur over a broad range of bias voltages, limiting the controllability of device characteristics. Here we show that similar zero field splitting can be achieved in the much more reliable quantum well geometry.

\begin{figure}[b]
  \includegraphics[width=0.45\textwidth]{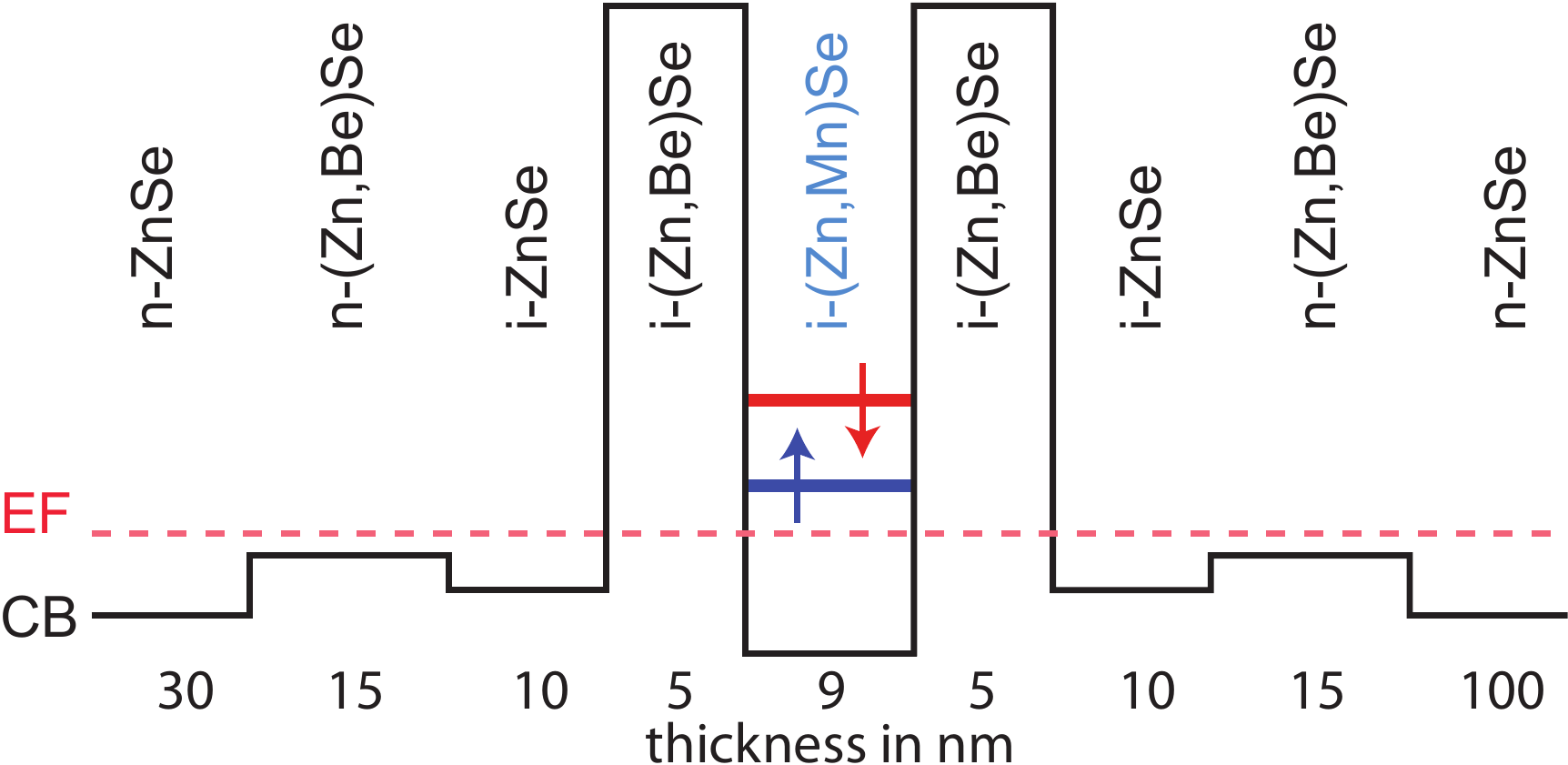}\\
  \caption{Schematic conduction band profile of the resonant tunneling diode at zero bias, with the spin degeneracy in the quantum well lifted in the quantum well.}
  \label{fig.cbprofile}
\end{figure}

We investigate an all-II-VI RTD grown on a GaAs substrate. The active RTD region contains a 9 nm Zn$_{0.96}$Mn$_{0.04}$Se quantum well layer sandwiched between two 5 nm Zn$_{0.7}$Be$_{0.3}$Se tunnel barriers. Proper contact layers are applied on each side of this structure to allow for measurements of transport through the layer stack (fig. \ref{fig.cbprofile}). The quantum well layer is made from a DMS that exhibits giant Zeeman splitting in an external magnetic field, which is described by a modified Brillouin function \cite{TWARDOWSKI1984,Slobodskyy2003} with a pair breaking contribution at high magnetic fields \cite{Shapira1984}. Lifting the degeneracy of the quantum well spin states with an external magnetic field allows the RTD to be used as a voltage controlled spin filter \cite{Slobodskyy2003}. The I-V characteristic shows current peaks at two different bias voltages as long as the splitting is large enough to resolve the separate spin up and down resonances.

\begin{figure}[tb]
  \includegraphics[width=0.45\textwidth]{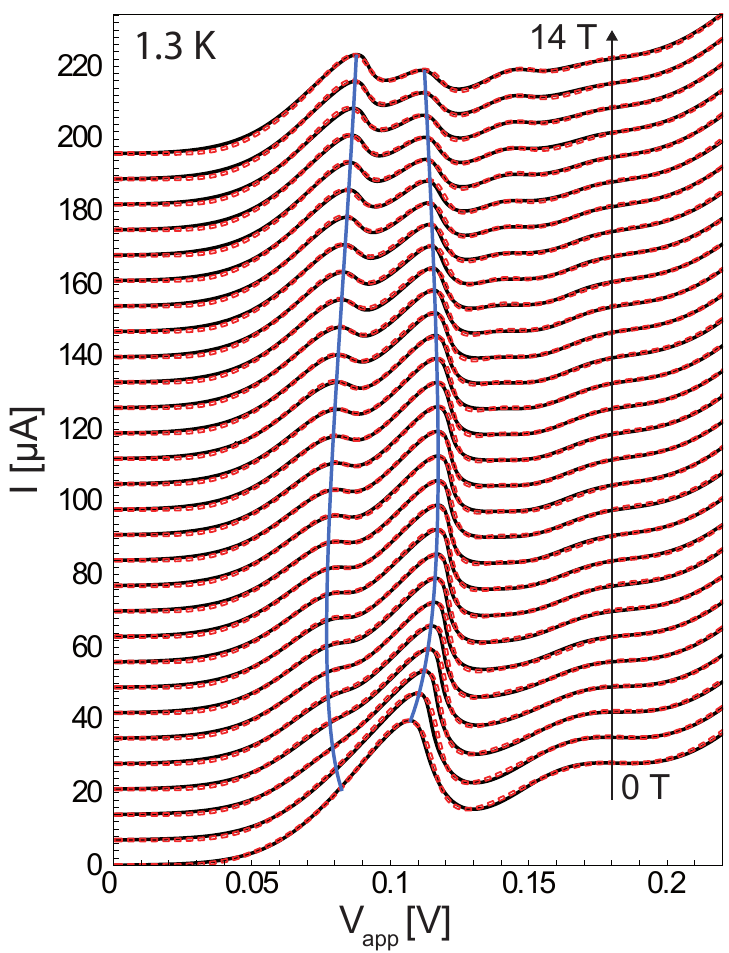}\\
  \caption{Fits (red dashed curves) to the I-V characteristics (black) at 1.3 K and at magnetic fields from 0 to 14 T, applied perpendicular to the layer stack. Each curve is offset by 14 $\mu$A on the current axis for clarity. The blue lines are a guide to the eye to emphasize the apparent peak splitting at B=0 T.}
  \label{fig.fit}
\end{figure}

The black lines in fig. \ref{fig.fit} show I-V characteristics for measurements at 1.3 K from 0 to 14 T. Similar results for fields up to 6 T have previously been successfully described \cite{Slobodskyy2003} using a model based on taking the conductance of a single spin channel to be one half of the B=0 T curve, applying Brillouin splitting to the quantum well levels and recombining the contribution of the two spin channels into a total I-V curve by using Kirchhoff's laws. Such a model implicitly assumes spin degeneracy at B=0 T, and obviously breaks down if that condition is not fulfilled. The data presented here, which include higher magnetic fields than available previously, suggest that a modified picture of the zero-field tunneling process is necessary.

As shown by the blue lines in fig. \ref{fig.fit} the data is suggestive of the peak splitting not vanishing at B=0 T. More importantly, the peak in the zero field I-V characteristic is also less symmetric than each of the split peaks at high magnetic fields, and the resonance in the zero field curve is much broader than that of the individual resonances in the 14 T curve. Both the asymmetry and the increased width of the peak in the B=0 T curve may be a consequence of this peak actually being comprised of two resonances occurring at somewhat different bias voltages. These considerations indicate the need for a different modeling scheme.

 \begin{figure}[tb]
  \includegraphics[width=0.45\textwidth]{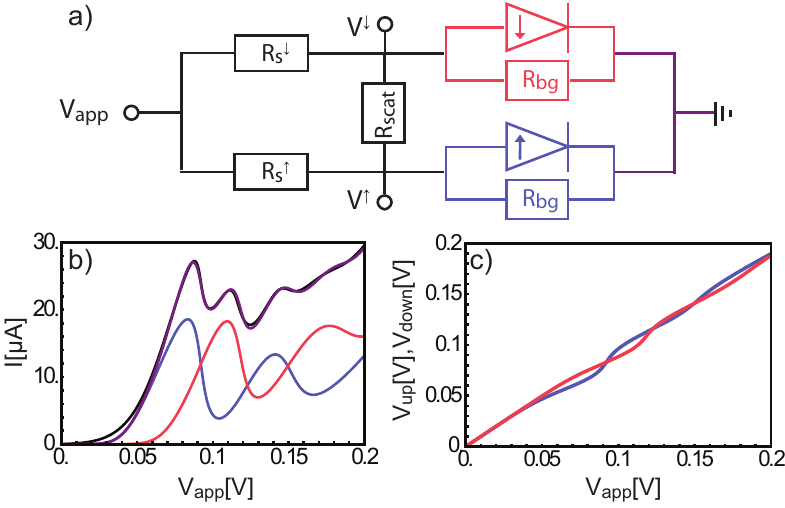}\\
  \caption{a) Resistor model for the two spin RTDs in parallel. b) Measurement and fit for the 14 T I-V characteristic. The blue and red curves represent the current carried by the spin up and down species, adding up to the purple curve which is the fit to the measurement (black curve). c) Plot of the potentials $V^\uparrow$ and $V^\downarrow$ at each spin diode as a function of the applied bias voltage.}
  \label{fig.kirchhoff}
\end{figure}

Bottom-up approaches to modeling such data have been reported \cite{Havu2005,Saffarzadeh2005}, but these typically treat an idealized system ignoring considerations such as contact resistances. In fig. \ref{fig.kirchhoff} we show that such considerations are important. Fig \ref{fig.kirchhoff}a gives the equivalent circuit of our real device in the two channel model, and includes magnetic field dependent contact resistances $R_{s}^{\uparrow,\downarrow}$, an interface scattering term $R_{scat}$ and a non resonant contribution to the tunneling current, $R_{bg}$. The active region of the RTD is represented by the two diodes, one for each spin channel, and each with a voltage and magnetic field dependent resistance $R_{bg}$ in parallel. While the diode carries the resonant part of the current including the LO-phonon replica contributions, the background current through $R_{bg}$ accounts for electrons tunneling off resonance through the double barrier region.

To obtain an expression for the highly non-linear resistance resulting from the resonant tunneling transport, one normally assumes a Lorentzian shaped transmission at the resonance condition. The peaks in our experiment show a Gaussian line shape with a bandwidth much broader than the expected injector Fermi energy (multiplied by the lever arm). Thus an additional broadening mechanism, probably stemming from imperfect interfaces at the active RTD region, dominates the resonance width \cite{EPAPS}. Local potential fluctuations, caused by well width fluctuations \cite{Zrenner1994} or inhomogeneous alloy or doping concentrations \cite{Makarovsky2010}, impose an additional in-plane confinement for tunneling electrons thus creating 0D type tunneling states, so called interface quantum dots \cite{Gammon1996}. Since our device is $100^2$ $\mu m^2$ and these fluctuations are typically on a nm scale, we sample over an ensemble of these states in our vertical transport measurements. One can view this configuration as a large number of 0D resonant tunneling diodes in parallel, each with its own resonance conditions. This results in a broadened Gaussian line shape \cite{Zrenner1994} for the overall resonant conductance feature. The LO-phonon replica are described by additionally broadened Gaussian conductance peaks with reduced amplitudes and an energetic separation from their respective spin-split resonance peaks of 31.7 meV, the LO phonon energy of bulk ZnSe \cite{Landolt-Boernstein}.

For the non-resonant background current we use a transfer matrix model consisting of two tilted barriers where the quantum well was omitted in order to remove resonant contributions. The potential drop over the quantum well region, which effectively lowers the second barrier, plays an important role, and is explicitly taken into account. The resulting transmission is proportional to the non-resonant tunneling of emitter electrons and fits well to the measurement at high bias voltage, where the contribution of resonant tunneling is small.

Due to the contact resistances $R_s^{\uparrow,\downarrow}$, the two voltage nodes $V_\uparrow$ and $V_\downarrow$ in fig. \ref{fig.kirchhoff}a are not necessarily at equipotential for a given applied bias voltage $V_{app}$. Fig. \ref{fig.kirchhoff}c shows the potential at the points $V_\uparrow$ and $V_\downarrow$ as a function of  $V_{app}$. When a resonance condition is reached for either of the spin diodes, the resistance of that spin diode drops and the potentials across each of the diodes is altered accordingly. While we have experimental access to $V_{app}$, the transport theory for resonant tunneling only describes the active region of the device. Thus considering the contact resistances is vital for fitting any RTD model to actual experiments. As an example, the resulting fits for a magnetic field of 14 T are presented in fig. \ref{fig.kirchhoff}b where contributions from both the spin up and down channels are shown as well as how they add up to produce a fit (purple curve) to the observed measurement (black curve). While the conductance of a resonant channel is perfectly symmetric on an energy scale, fig. \ref{fig.kirchhoff}b+c show how in a real device, the contact resistances influence the shape of the resulting I-$V_{app}$ characteristics.

As is clear from the circuit diagram in fig. \ref{fig.kirchhoff}a, the total current traversing the device is given by

\begin{eqnarray}\label{eq.current}
    &I(V_{app})=&V^\uparrow\left(\sigma^\uparrow(V^\uparrow)+\sigma_{LO}^\uparrow(V^\uparrow)+\sigma_{bg}^\uparrow(V^\uparrow)\right)+\nonumber\\
    &&V^\downarrow\left(\sigma^\downarrow(V^\downarrow)+\sigma_{LO}^\downarrow(V^\downarrow)+\sigma_{bg}^\downarrow(V^\downarrow)\right)\\
    &\hspace{-24pt}\mbox{with }&\hspace{-12pt}\sigma^{\uparrow,\downarrow}(V^{\uparrow,\downarrow})\propto p^{\uparrow,\downarrow}\cdot \exp\left(\frac{(l (E-E_0^{\uparrow,\downarrow}))^2}{2  {\Gamma^{\uparrow,\downarrow}}^2}\right)\label{eq.narf}
\end{eqnarray}

where $\sigma^{\uparrow,\downarrow}$, $\sigma_{bg}^{\uparrow,\downarrow}$ and $\sigma_{LO}^{\uparrow,\downarrow}$ are the conductances for the spin channels, the background contributions and the LO-phonon replica peaks respectively. $l$ is the lever arm linking the energy scale in the quantum well to the diode bias voltages $V^{\uparrow,\downarrow}$, $E_0^{\uparrow,\downarrow}$ is the energy between the spin levels and the conduction band edge, $p^{\uparrow,\downarrow}$ are fitting parameters for the amplitudes of the spin conductances (and thus yield the spin polarization) and $\Gamma^{\uparrow,\downarrow}$ are the variances of the Gaussians describing the energy level distribution for the spin channels. Equation \ref{eq.narf} is also used for $\sigma_{LO}^{\uparrow\downarrow}$ but with different variances $\Gamma_{LO}^{\uparrow\downarrow}$, amplitudes $p_{LO}^{\uparrow\downarrow}$ and $E_{0,LO}^{\uparrow\downarrow}=E_{0}^{\uparrow\downarrow}+31.7$ meV.

Our detailed model therefore consists of solving the equivalent circuit of fig. \ref{fig.kirchhoff}a for an RTD with a spin split resonance and the associate LO-phonon replica. Since the zero field I-V characteristic is a superposition of two strongly overlapping peaks, the best starting point for the fits is the high magnetic field data, where one easily can find the proper variances $\Gamma^{\uparrow,\downarrow}$ and $\Gamma_{LO}^{\uparrow\downarrow}$ of the resonant peaks and LO-phonon replicas. Starting at 14 T, the I-V characteristic for each magnetic field is fitted by adjusting $p^{\uparrow,\downarrow}$, $p_{LO}^{\uparrow\downarrow}$ and $E_0^{\uparrow,\downarrow}$. We also allow for a magneto-resistance effect in the contacts $R_s^{\uparrow,\downarrow}$ and in the scattering channel $R_{scat}$. By including a magnetic field dependence of $\Gamma_{LO}^{\uparrow\downarrow}$, we account for the small, experimentally observed field dependent broadening of the replica peaks.

The resulting fits are shown as red dashed lines on top of the I-V characteristics in fig. \ref{fig.fit}, while in fig. \ref{fig.kirchhoff}b the contributions of spin-up and spin-down electrons to the 14 T I-V characteristic are illustrated.
One would a priori expect a Brillouin function to describe the magnetic field dependence of the splitting \cite{Slobodskyy2003}. The measurements shown in fig. \ref{fig.fit} exhibit a very different behavior. At low magnetic fields we observe that instead of a spin degeneracy, the I-$V_{app}$ characteristic is properly fit only by allowing for finite splitting even at zero magnetic field \cite{EPAPS}. We have previously observed such a remanent zero field splitting in the zero dimensional resonant tunneling states of self assembled CdSe quantum dots \cite{Gould2006}. Here the quantum well is nominally a two dimensional object. As previously discussed, however, various inhomogeneities cause the current transport to be effectively mediated by a large ensemble of parallel paths each flowing in a local environment. The relatively low number of magnetic atoms influenced by each of these regions means that each will statistically have, on average, a net magnetization at zero field \cite{Dietl1982}. This effect is further enhanced by the presence of the spin of the tunneling electron \cite{Dietl1983}.

The energy separation between spin-up and spin-down peaks is 15 meV at B=0 T as determined by the fit. This energy is not necessarily the same as the splitting of the two spin states. As the measurement is always referred to the conduction band of the emitter, this energy difference is influenced by the different bias conditions needed to align each spin state to the emitter.

The first peak at lower bias voltage is suppressed while the second peak is enhanced in the B=0 T I-$V_{app}$ characteristic. For a small energetic splitting in the resonant state, one would expect similar conductances for the two transport channels. A change in the confinement caused by the splitting will influence the amount of leakage of the quantum well wave function into the emitter, while for each $V_{app}$ the resulting change in symmetry of the double barrier will affect the transmission \cite{LURYI1989}. A higher bias voltage will also drive more current at the same conductance. From transfer matrix calculations for the transmission probabilities of the double barrier we conclude that different biasing conditions alone cannot explain the magnitude of the effect on the amplitudes of the B=0 T spin currents.

While the peak positions stay constant at intermediate fields because the Brillouin function saturates, above 8 T there is a clear reduction in the splitting of the peaks on the bias voltage axis. A reason for this reduction is likely the Zeeman splitting of the emitter electrons, since both ZnSe and (Zn,Mn)Se have a positive g factor and the resulting splitting $\Delta V_{res}$ on the voltage axis is given by $\Delta V_{res}=l \left(g_{QW}-g_{E}\right)\mu_B B$, where $l$ is the lever arm of the device and $g_{QW}$ and $g_{E}$ are the effective g-factors of the ZnSe emitter and the quantum well electron states respectively. From the fits we obtain a slope of 0.47 meV/T (-0.18 meV/T) for the spin-up (down) peak. The corresponding g-factors are $g_{\uparrow}=16.3$ and $g_{\downarrow}=7.3$, far greater than the bulk ZnSe value of 1.1 \cite{Landolt-Boernstein}. Possible explanations for this increase in magnetic response include that tunneling electrons at the interface to the barrier cannot be treated in the free electron picture of a parabolic s-type conduction band, that there is a dilute Mn concentration in the emitter due to diffusion during growth, or that the energetic distance to the resonant quantum well state is altered by spin selective band bending of the emitter. The peak amplitudes are also strongly magnetic field dependent. The asymmetry of the effective g factors for the emitter polarization suggests an effect that is linked to the resonance bias conditions. The two peaks occur at different bias voltages and therefore have different conduction band bending conditions. This bending changes the number of available electronic states for resonant tunneling, strongly influencing tunneling currents. This factor can easily surpass the effect of Zeeman splitting. The resulting effective g-factors are therefore not purely a result of the electron spin interacting with the magnetic field but also of the feedback mechanisms induced by changes in the potential landscape \cite{EPAPS}.
Different transmittances of the spin channels may also result in spin sensitive charge build-up in front of the barrier that can influence not only the amount of available states in front of the barrier, but also the bias voltage needed to attain the resonance conditions \cite{Choi1992}.

\begin{figure}[tb]
  \includegraphics[width=0.45\textwidth]{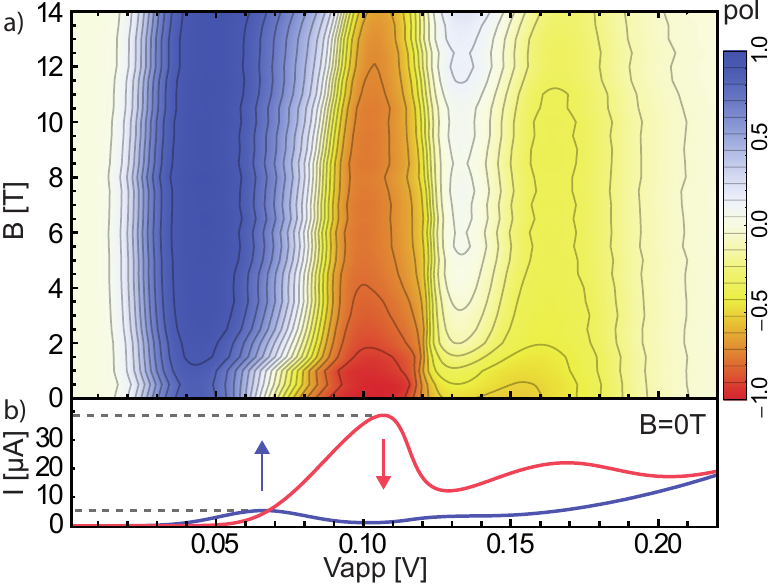}\\
  \vspace{-4pt}
  \caption{Current spin polarization as a function of applied bias voltage and magnetic field.}
  \label{pol.fit}
\end{figure}

The amplitudes $p^{\uparrow\downarrow}$ we obtain from the above fitting process give quantitative results for the spin polarized currents. To use this device as a detector for the emitter spin polarization one would need to link the emitter polarization to the amplitude of the traversing spin currents. The increased and asymmetric magnetic feedback that is evidenced by the movement of the spin peaks suggests that other effects in addition to pure Zeeman splitting of the emitter are involved \cite{EPAPS}. Therefore usage as a detector for the emitter spin polarization is difficult. Within our model it is possible to evaluate currents for spin-up and down electrons separately, thus allowing for a quantitative analysis of the polarization of the current traversing the device. Fig \ref{pol.fit}a) shows the current spin polarization as a function of magnetic field and bias voltage. Blue (red) indicates a spin up (down) polarization of the current. The non resonant background current is not spin selective and therefore the current polarization $P_c$ plotted in fig \ref{pol.fit} is given by

\begin{equation}\label{eq.pol}
P_c=\frac{I_\uparrow-I_\downarrow}{I_\uparrow+I_\downarrow+I_{bg(\uparrow+\downarrow)}}
\end{equation}

The splitting of the spin levels in the external magnetic field and the changes in the amplitudes of the resonant peaks lead to a polarization above 90\% for the spin-up peak at B=14 T, while the spin-down peak polarization decreases to below 60\%. While a high degree of polarization of both spin types can be achieved at all measured magnetic fields, counterintuitively, despite the paramagnetic nature of bulk (Zn,Mn)Se, the maximum polarization efficiency is achieved without applying an external magnetic field, where 80\% for spin-up and 90\% for spin-down is observed, as evidenced by the I-V curve of the two channels for B=0 T presented in fig. \ref{pol.fit}b). Similar results for a second device and at various temperatures are presented in \cite{EPAPS}.

In summary, we have shown high spin polarizations can be achieved due to formation of BMP like states in the active RTD region. The resulting microscopic magnetization for the tunneling electrons lifts the spin degeneracy and provides two separate transport channels. Feedback mechanisms stemming from the influence of different biasing conditions both increase the energy splitting of the peaks and influence their amplitudes, resulting in high degrees of current spin polarization. Our model allows for good fits to the device characteristics and thus quantitative analysis of the polarization. Not only does this model confirm the findings of reference \cite{Slobodskyy2003} that the device can work as a voltage controlled spin filter at moderate magnetic fields, but it also establishes that the local spin polarization efficiency not only remains, but is even enhanced in the absence of a magnetic field.

\begin{acknowledgments}
We thank A. Slobodskyy and T. Slobodskyy for sample preparation and measurements, and the DFG SPP 1285 "Semiconductor Spintronics" for financial support.
\end{acknowledgments}

\cleardoublepage
\begin{appendix}
\chapter{\large{\textbf{supplementary data}}}

\begin{figure}[htb]
  \includegraphics[width=0.45\textwidth]{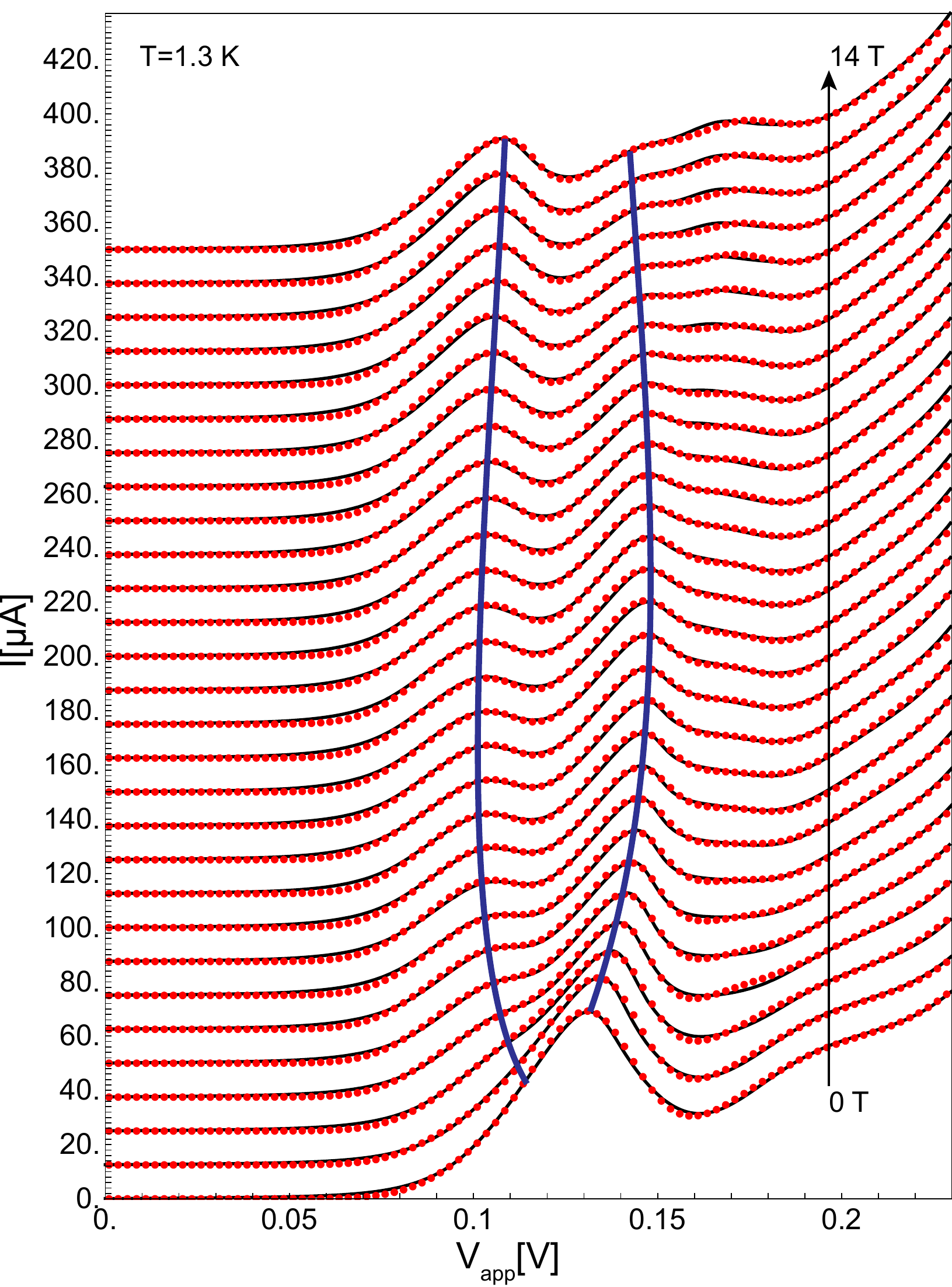}\\
  \caption{I-V characteristics (black lines) and fits (red dots) of a sample with 8\% Mn and 6\% thinner layers at magnetic fields from 0 to 14 T (in steps of 0.5 T) and T=1.3 K. Each curve is offset by 14 $\mu$A on the current axis for clarity.}
  \label{fig:fit_overview}
\end{figure}

Fig. \ref{fig:fit_overview} shows a comparison of our model to experimental data for a second type of sample. This device has 8\% Mn instead of 4\%, and all layers are 6\% thinner that the device in the paper. The graph presents I-V characteristics at 1.3 K for magnetic fields from 0 to 14 T (black lines). The fits (red dots) again agree well with the experiment, even though, due the high Mn content, the spin down resonance is merged with the spin up replica peak. All fitting parameters show the same magnetic field dependence as in the manuscript. As in the text, the blue lines are a guide to the eye.

\begin{figure}[htb]
  \includegraphics[width=0.45\textwidth]{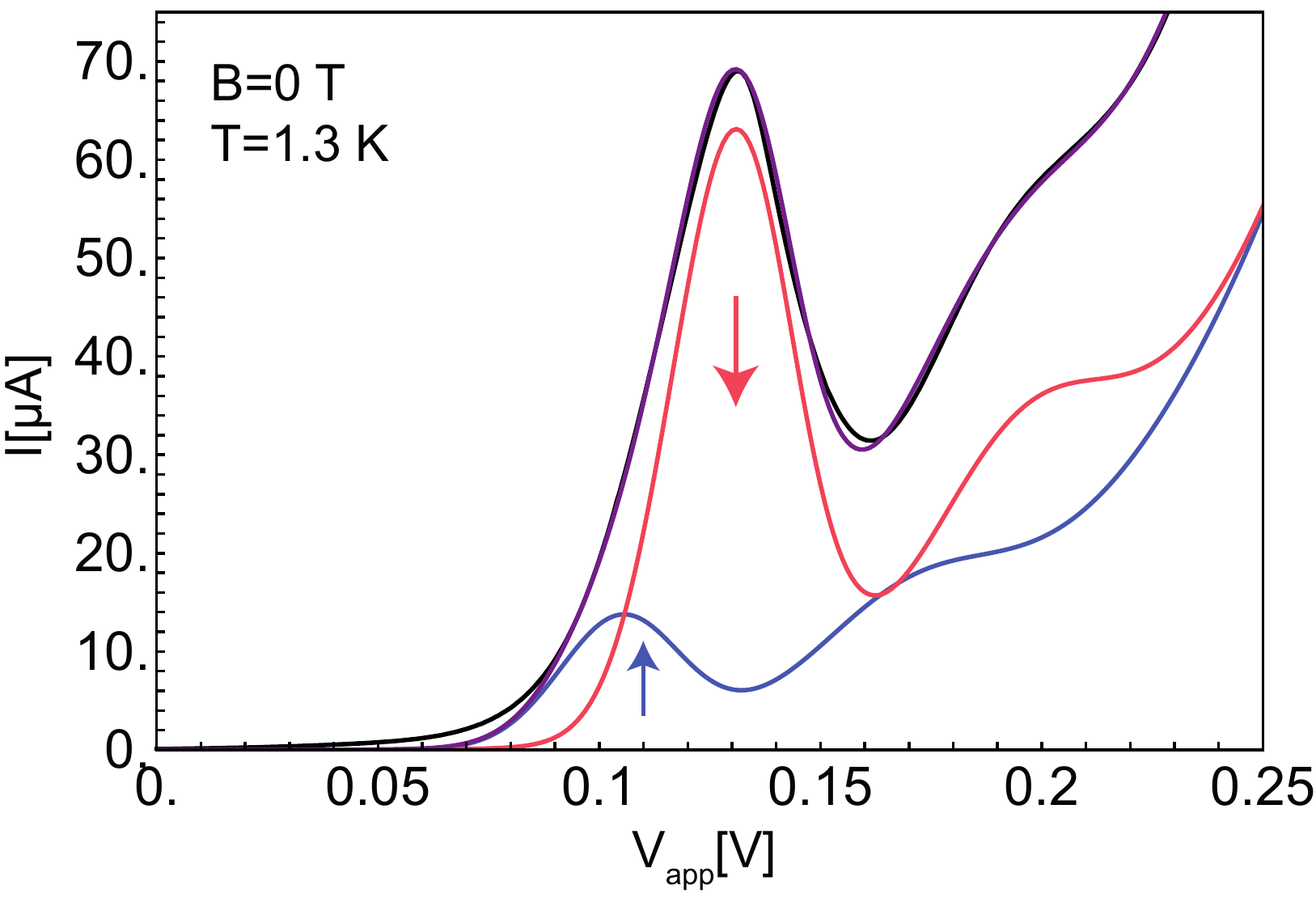}\\
  \caption{Also the control sample is highly suggestive of zero field splitting. The I-V characteristic (black line) is fitted (purple line) by adding the spin up (blue line) and down (red line) currents.}
  \label{fig:fit_0T}
\end{figure}

\begin{figure}[htb]
  \includegraphics[width=0.45\textwidth]{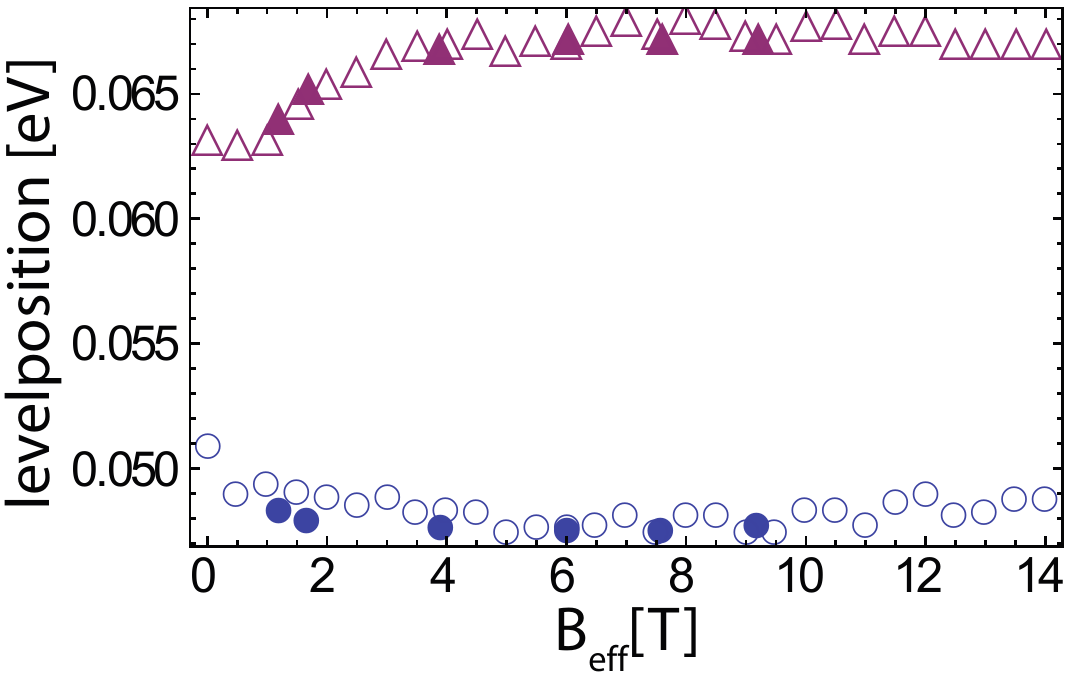}\\
  \caption{Energetic distance to the conduction band of spin up (circles) and down (triangles) levels in the quantum well. Filled symbols show results from fits at various temperatures at B=6 T, while empty symbols show fits at various magnetic fields at T=1.3 K.}
  \label{fig:levelpos}
\end{figure}

Fig. \ref{fig:fit_0T} gives the details of the fit to the zero magnetic field I-V characteristic, again showing zero field splitting with a spin up peak of reduced amplitude and a pronounced spin down peak. In this case, a spin polarization of 80\% (92\%) for spin up (down) electrons is achieved.
We now consider the effect of temperature dependence, by analyzing data taken at 6 T for various temperatures ranging from 45 mK to 15 K. For each temperature Temp one can solve the equation $$\mbox{Brillouin[6 T,Temp]}=\mbox{Brillouin[B}_{eff}\mbox{,1.3 K]}$$ to determine at which magnetic field B$_{eff}$ a curve from the 1.3 K dataset has the same level splitting in the quantum well as the 6 T curve at the given temperature. Fig. \ref{fig:levelpos} presents the level positions of the resonant spin states for both the 1.3 K data set of fig. \ref{fig:fit_overview} and this temperature dependent measurement. The open symbols are for the 1.3 K dataset, and the x-axis is then directly the magnetic field at which the measurement was performed. The solid symbols are for the temperature dependent data, plotted against B$_{eff}$ as described above. This comparison confirms that the movement of the peak position is a result of changes in the band diagram, and not a result of any deformation of peak shape, as these would not be stable under the different environmental conditions.

\begin{figure}[htb]
  \includegraphics[width=0.45\textwidth]{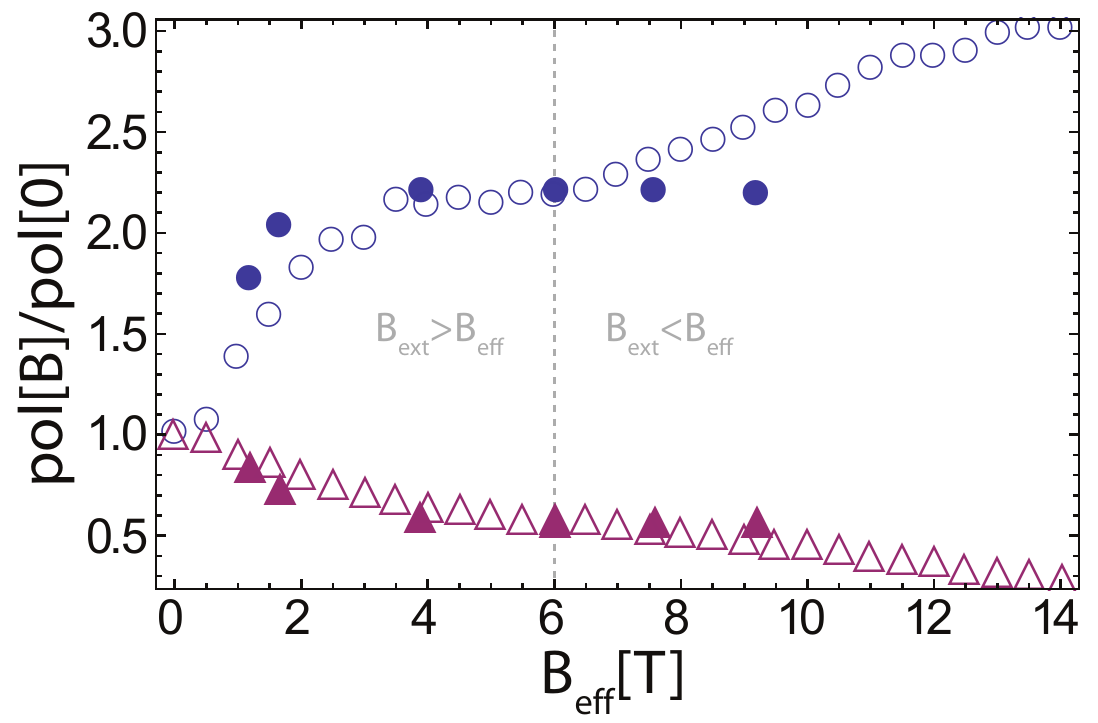}\\
  \caption{Change in amplitudes of the spin up (circles) and down (triangles) levels with applied external magnetic field (normalized to the B=0 T amplitudes). Filled symbols show results from fits at various temperatures at B=6 T, while empty symbols show fits at various magnetic fields at T=1.3 K. }
  \label{fig:pol}
\end{figure}

\begin{figure}[htb]
  \includegraphics[width=0.45\textwidth]{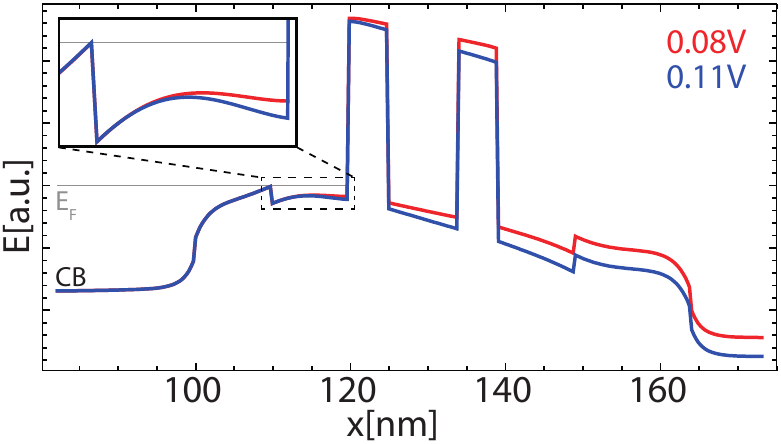}\\
  \caption{Self consistent conduction band profile of the resonant tunneling diode at the approximate resonance biases. The inset shows the increase of tunneling states due to the applied bias voltage.}
  \label{fig:cb_profile}
\end{figure}

\begin{figure}[htb]
  \includegraphics[width=0.45\textwidth]{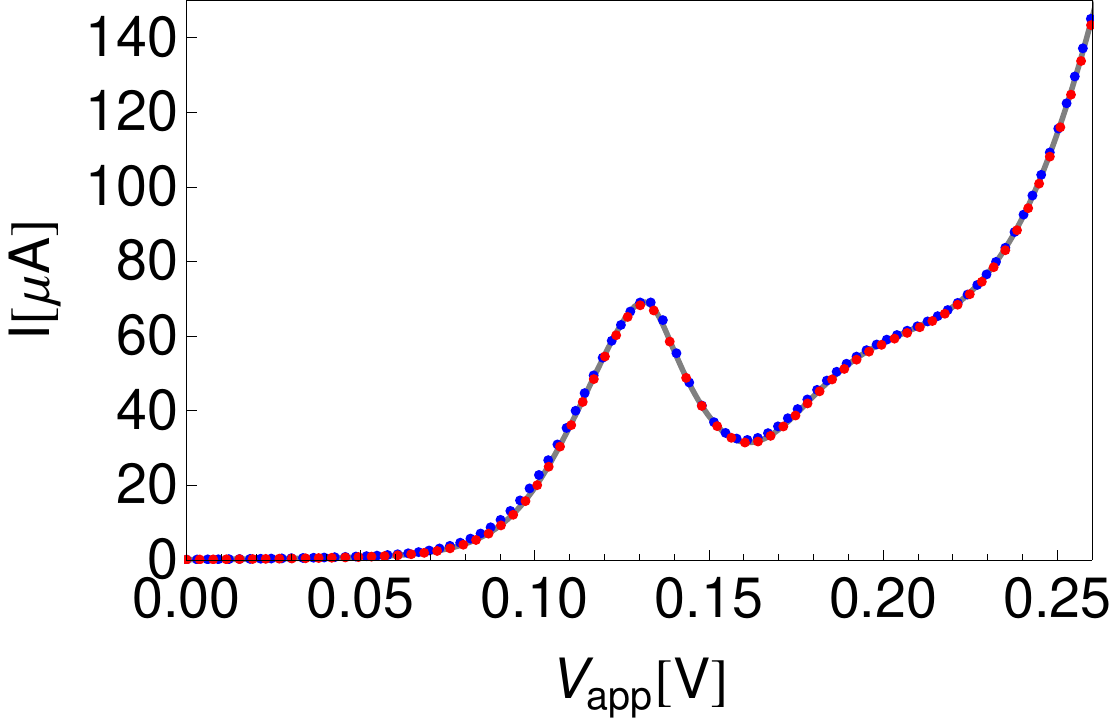}\\
  \caption{Comparison of I-V characteristics at T=40 mK, 1.3 K and 15 K. The resonance does not sharpen at low temperatures.}
  \label{fig:tempcompare}
\end{figure}

A similar analysis also is conducted for the amplitudes of the spin channel conductances in fig. \ref{fig:pol}. The polarization values acquired from fits of measurements at various temperatures differ from those acquired from the magnetic field dependent measurements at T=1.3 K. After correcting for the level splitting using B$_{eff}$, the only visible difference between the two sets of measurements are peak amplitudes. This is because a constant quantum well splitting is maintained in the two configurations, which then only differ in emitter polarization due to the applied external magnetic field. Fig. \ref{fig:pol} shows that indeed, the only field where both polarization values are identical, is at B$_{eff}$= 6 T, where both temperatures are the same. For B$_{eff}<$ 6 T the temperature measurements show higher polarization since, while the splitting in the quantum well is maintained constant, the external magnetic fields effect on polarizing the emitter produces a higher polarization of the spin current. The opposite is true for B$_{eff}>$ 6 T. The results of fig. \ref{fig:pol} thus suggest that both the splitting in the quantum well and the polarization of the emitter influence the spin polarization of the resonant current.

As mentioned in the text, the reason why a change in quantum well splitting already results in a change of current polarization is the change in the bias voltage needed to reach resonant conditions for each of the spin channels. Fig. \ref{fig:cb_profile} shows self-consistent calculations of the conduction band profile at the resonance condition for the 4\% Mn sample. As the inset shows, the Fermi energies differ by approximately 20\% for the two resonance conditions. Since the maximum current flows when the quantum well level is aligned with the conduction band edge and is proportional to the cross-sectional plane $A=\pi k_F^2$ of the emitter Fermi sphere at constant E$_z$ this would result in a $\approx$44\% change of the peak amplitudes.

Lastly, in fig. \ref{fig:tempcompare} we plot the B=0 T I-V characteristics of 40 mK (red dots), 1.3 K (gray line) and 15 K (blue dots), showing that in the absence of magnetic field, temperature does not have any influence on the I-V characteristic. This indicates that a much stronger broadening mechanism is at work in the device; namely the potential fluctuations at the quantum well interface.

\end{appendix}

\end{document}